\input epsf
\newcommand{\myref}[1]{(\ref{#1})}

\def\d{{\rm d}}
\def\beq{\begin{equation}}
\def\eeq{\end{equation}}
\def\bey{\begin{eqnarray}}
\def\eey{\end{eqnarray}}

\documentstyle[twocolumn]{mn}
\begin{document}

\title[Can galactic nuclei be non-axisymmetric?]
      {Can galactic nuclei be non-axisymmetric? --- The parameter space 
       of power-law discs}
\author[Zhao et al.]
       {HongSheng Zhao, C.\ Marcella Carollo \& P.\ Tim de Zeeuw
       \thanks{E-mail: hsz@strw.leidenuniv.nl, 
                       marci@pha.jhu.edu, \hfill\break
                       \null\hskip 1.36truecm tim@strw.leidenuniv.nl} \\
       Sterrewacht Leiden, Niels Bohrweg 2, 2333 CA, Leiden, The Netherlands\\
       Johns Hopkins University, 3701 San Martin Drive, Baltimore MD 21218, 
                               USA}
\date{Accepted $\ldots$ 
      Received $\ldots$;
      in original form $\ldots$}

\maketitle
\label{firstpage}

\begin{abstract}
The shape of a cusped galactic nucleus is constrained by the range of
shapes of orbits in its gravitational potential.  It is shown for
scale-free non-axisymmetric discs that while a plausible elongated
density model requires at least some orbits to spend more time near
the major axis than anywhere else, both regular boxlets and tube
orbits generally cross the major axis too fast for
self-consistency.  If galaxies host inner nuclear discs or flat bars
with a cuspy surface light profile ($\gamma=|{\d \log \mu / \d \log
r}|>0$), their ellipticity $1-{b / a}$ cannot be greater than about
${\gamma / 2}$. Discs or bars with a shallow central profile ($\gamma
\le 0.3$) should not be strongly elliptical.
\end{abstract}


\section{Introduction}

Current formation theories emphasize the roles of dissipation and
galaxy interaction as major processes in shaping present day
galaxies, but is there any stringent limit on the shapes and density
profiles of galaxies from general conditions such as equilibrium and
stability alone?  Particularly, are there stable triaxial equilibria
with realistic radial density profiles?  This has been an out-standing
stellar dynamical question ever since Binney (1978) invoked triaxial
equilibria to account for the observed flattened shape of elliptical
galaxies and their lack of rotation.  Theoretically triaxial
equilibria exist at least for systems with a finite density core and a
wealth of box orbits and tube orbits. This was demonstrated for
general models by Schwarzschild (1979, 1982), and for models with
separable potentials by Statler (1987).  However, the traditional
assumption of a finite core in every galaxy was challenged by recent
observations of nuclei of nearby elliptical galaxies with the Hubble
Space Telescope. It was found that giant ellipticals have a power-law
surface density distribution $\mu
\propto r^{-\gamma}$ with $0 < \gamma < 0.3$ near the center; the
slope steepens to as much as 0.3--1.3 for small ellipticals, with the
dividing line at $M_B \approx -20$ mag (Crane et al.\ 1993; Jaffe et
al.\ 1994; Lauer et al.\ 1995; Carollo et al.\ 1997; Faber et al.\
1997).  These observations call for a re-examination of the existence
of triaxial equilibria in a potential with a divergent force ($F
\propto \mu \rightarrow \infty$ for $\gamma>0$) at the center (Gerhard
\& Binney 1985; de Zeeuw \& Carollo 1996).  Whether triaxial models of
this kind exist also becomes a key uncertainty in, e.g., interpreting
the kinematic and photometric data of galactic nuclei, weighing their
central black holes and reconstructing the formation history of these
systems.

The conclusion that at least some strongly non-axi\-sym\-me\-tric
models with a steep central cusp are probably not in rigorous steady
state is based on a handful of three-dimensional dynamical models.
These have been built with various implementations of Schwarzschild's
method, in which individual orbits are populated so as to match the
model density, and include three-dimensional scale-free models with
logarithmic potentials (Schwarzschild 1993), and two non-scale-free
models (Merritt \& Fridman 1996). Unfortunately, the power of these
few numerical experiments is limited when it comes to exploring the
parameter space. It is not clear how to extrapolate results obtained
for a few models to a general statement about the whole class of
cusped triaxial potentials, because the meaning of a small mismatch in
the reconstructed density, which is often of the order of one percent
or less, has to be interpreted on a case-by-case basis.  Kuijken
(1993, hereafter K93) showed in his systematic study of
two-dimensional non-axisymmetric models with a logarithmic potential
that whether the numerically constructed model is in equilibrium
depends sensitively on numerical details, including the resolution of
the spatial grid for the mass model, the grid for orbital initial
conditions, the number of orbits used, and the integration time for
each orbit. Although it may be feasible with present-day computer
technology to carry out a massive numerical search in the
multi-dimensional parameter space (axis ratios, inner and outer
density slopes with the possible addition of a central black hole and
the tumbling speed of the potential), it is intrinsically difficult in
this approach to pinpoint the exact origin of any mismatch between the
orbits and the density model. For example it has not been
well-understood why replacing box orbits in a cored potential by
boxlets in a cusped potential upsets the equilibria.\looseness=-2

In this paper we present a new approach to study self-consistency of a
general non-axisymmetric model. We restrict ourselves to the
two-dimensional case of scale-free discs, and show that a non-trivial
and necessary condition for the existence of a self-consistent
elongated disc is that the angular speed of the regular boxlet and
tube orbits when they cross the major axis should be consistent with
the local curvature of the density distribution.  It is well known
that boxlets are less useful than box orbits when it comes to fit the
model density near the major axis because the boxlets have their
density maxima at the turning points rather than on the major axis
(e.g., Pfenniger \& de Zeeuw 1989; Schwarzschild 1993).  K93 made the
interesting observation that a box orbit has only one `corner' per
quadrant, while a boxlet orbit has two or more correlated `corners'
per quadrant, which makes them less flexible in fitting the model
density.  He also suggested that the lack of self-consistency in
elliptical models is likely due to the spiky angular distribution of
boxlets rather than to the lack of flattened boxlets.  Syer \& Zhao
(1998, hereafter SZ98) suggested that the self-consistency is first
broken down near the symmetry axes of the model.  Unfortunately, the
result of SZ98 is limited to a special subset of the separable
non-axisymmetric potentials introduced by Sridhar \& Touma (1997, 
hereafter ST97). 

The structure of this paper is as follows. \S 2 gives a rigorous
formulation of the problem for scale-free discs. \S 3 illustrates the
requirement on the curvature of the model with specific orbits in
elliptic disc potentials.  \S 4 gives the results of fitting the
curvature, and \S 5 examines the assumptions in the model, and
discusses generalizations of the method to three-dimensional and
non-scale-free systems, as well as the implications for barred
galaxies.

\section{Scale-free non-axisymmetric discs}

\subsection{Formulation}
Consider a general two-dimensional scale-free non-axi\-sym\-me\-tric
disc potential
\beq\label{pot}
\phi(r,\theta) \propto r^{\alpha} p(\theta), \qquad {-1 \le \alpha \le 1},
\eeq
where $p(\theta)$ defines the angular shape, with $\theta=0$ and $\pi/2$
the directions of the minor and major axis, respectively. We consider 
surface densities of the form
\beq
\mu(r,\theta) \propto r^{-\gamma} s(\theta), \qquad 
                                    0 \le \gamma < 2,
\eeq
where $\gamma$ is the cusp strength and $s(\theta)$ the angular shape
of the surface density. In self-consistent systems $s(\theta)$ and
$p(\theta)$ are related, and $\gamma=1-\alpha$.

The regular orbits in this potential can be grouped according to their
shapes (bananas, fishes, tubes, etc.; see Miralda--Escud\'e \&
Schwarzschild 1989; K93), irrespective of their sizes, into
self-similar families.  Each family can be characterized by a
dimensionless second integral, say, $I$, and the whole family can be
then built by rescaling one reference orbit with a trajectory
described by the polar coordinates $(r_I(t),\theta_I(t))$. The weights
among the self-similar `cousin' orbits should be prescribed in a
scale-free fashion such that each family produces a $r^{\alpha-1}$
power-law density distribution, as for the model density.  The
different families should be weighed according to a to-be-determined
positive function $w(I)\ge 0$ so as to reproduce the angular part of
the model density.

A clear discussion of the algorithm for constructing self-similar
non-axisymmetric models was given by Richstone (1980) and
Schwarzschild (1993) for three-dimensional models, and by K93 for
two-dimensional models.  These previous studies were focused
exclusively on the logarithmic potential, and it appears that no one
has applied the algorithm to other non-axisymmetric power-law
potentials, which are most relevant for galactic nuclei. Following
K93, we first divide the angular coordinate into $n$ slices with an
interval $\Delta \theta = {2 \pi /n}$, and $n$ approaching
infinity.  We then compare the amount of mass a reference orbit
deposits in each angular sector to the amount of mass required by the
density model in the same sector within the radius of the orbit.  If a
reference orbit $(r_I(t),\theta_I(t))$ spends a fraction $\Delta
(t/T)_{I,j}$ of the integration time $T$ in the $j$-th sector, then it
deposits an amount proportional to $w(I) \Delta (t/T)_{I,j}$.  By
comparison, the amount of mass prescribed by the density model in the
same angular sector within the radius $r_I(t)$ is,
\bey\label{mm}
\Delta m_{I,j} &\equiv &\Delta \theta 
             \int_{0}^{r_I(t)} \mu(r_1,\theta) r_1 dr_1 \\
               &\propto &  r^2_I(t) \mu(r_I(t),\theta) \Delta \theta.
\eey
To reproduce the model density using regular orbits alone we require that
\beq\label{dd}
\int dI w(I) 
\left<{ \Delta (t/ T)_{I,j} \over 
         \Delta m_{I,j} }\right> = 1, \qquad j=1,\ldots,n
\eeq
where we have taken time averages (as indicated by the brackets) of
all the times when an orbit comes to the same angle, and have summed
up all regular families by integrating the second integral $I$.

It is straightforward to verify this result for the scale-free
logarithmic disc of K93.  In this special case $\mu \propto r^{-1}
s(\theta)$, so $\Delta m_{I,j} \propto r_I(t) s(\theta) \Delta
\theta$.  Eq.\ \myref{dd} can then be rewritten in a form similar to K93,
$\int dI w(I) \langle \Delta(t/T)_{I,j} / r_I(t) \rangle \propto
s(\theta) \Delta \theta $.

\begin{figure} 
\epsfxsize=8.4truecm 
\centerline{\epsfbox{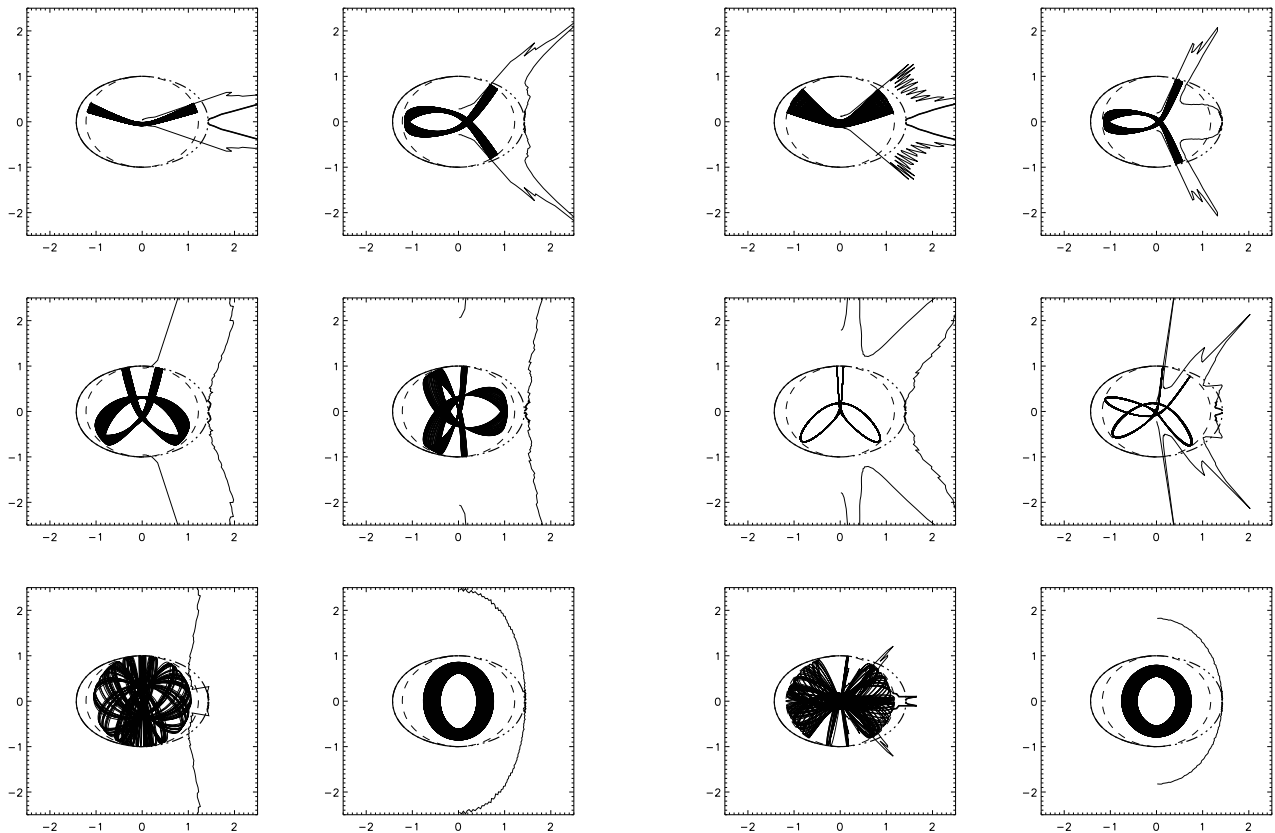}}
\caption{Various reference orbits (thin solid lines) with the same
energy in a two-dimensional scale-free elliptic disc with a surface
density power-law slope $\gamma =1-\alpha=0.8$ and axis ratio $q=0.7$.
The orbits shown are a regular banana orbit, a fish orbit, two high
resonance orbits, a stochastic orbit and a tube orbit.  Compare the
shape of the orbits with the shapes of the overplotted zero-velocity
curve (thin dashed line), the model density contour (dash-dotted line)
and the contour made by summing up the spatial distribution of the
`cousin' orbits, the scaled and reflected copy orbits (the outer solid
lines).  Near the major axis the contours of these orbits bend in the
opposite direction of the elliptical contours of the model
density. The similar contours for tubes bend in the right direction,
but not strong enough.}
\label{a2q7.ps}
\end{figure}

\subsection{Curvature at the major axis}

For the time being we assume that the angular momentum of a regular
orbit is always non-zero everywhere in the orbit (which is clearly not
the case near the turning point of a boxlet orbit), so that
$\dot{\theta_I} \ne 0$.  Taking the limit $\Delta \theta
\rightarrow 0$, we have
\beq
{\Delta t \over \Delta \theta} \rightarrow \|\dot{\theta}_I(t)\|^{-1},
\eeq
and upon substitution in eq.\ \myref{dd} and \myref{mm}, we obtain 
\beq\label{basic}
\int dI w(I) \left<\Gamma\right> = {\rm const}, 
\qquad
\Gamma \equiv { 1 \over \mu(r, \theta) \|J\|},
\eeq
where the weights $w(I)$ are non-negative and $J \equiv
r^2\dot{\theta}$ is the angular momentum.

To examine how the orbits fit the curvature of the density near the
major axis, we take the double derivative of eq.\ \myref{basic} with
respect to $\theta$, and substitute the result in the equations of
motion
\beq\label{eom}
\ddot r  = {J^2 \over r^3} - {\partial \phi \over \partial r},
\qquad
\dot J = - {\partial \phi \over \partial \theta}.
\eeq
The angular momentum $J$ is nearly constant in the vicinity of the
major axis where the torque $ -\partial \phi /\partial \theta
\rightarrow 0$, because the force is radial at the symmetry axes.
Evaluating the angular derivatives at the major axis of the potential
($\theta=\pi/2$), we obtain the following simple expression (see
Appendix A for a derivation):
\beq \label{Heq}
\int dI w(I) \left<
                 [ q^{-2}_\phi-(1+\gamma -\gamma \lambda)] 
                   {\Gamma \over K_\theta }\right> =0,
\eeq
where
\beq\label{lambda}
\lambda \equiv (1+\gamma) K_r + q^{-2}_\mu K_\theta
\vert_{\theta={\pi \over 2}} >0,
\eeq
and we have written $K_r \equiv {\dot{r}^2 \over r\partial_r \phi}$
and $K_\theta \equiv {r^2 \dot{\theta}^2 \over r\partial_r \phi}$
as the radial and tangential parts of
the kinetic energy scaled by the virial $r\partial_r\phi$.  The
quantities $q^{-2}_\phi$ and $q^{-2}_\mu$, defined as
\bey\label{qdef}
q^{-2}_X \equiv 1 + {\partial_\theta^2 X\over r \partial_r X} 
                \vert_{\theta={\pi \over 2}}, \qquad 
X=[\phi(r,\theta),\mu(r,\theta)] 
\eey
describe the curvatures of the equal potential and equal density
contours at the major axis (alternative expressions in terms of double
derivatives of $s(\theta)$ and $p(\theta)$ are given by eqs
\ref{aqmu} and~\ref{aqphi}).  Here $q_{[\phi,\mu]}$ are the axis
ratios of the best-fitting ellipses to the potential, respectively
density, contours near the major axis.  A similar result can be
derived for any angle $\theta$, but the expression is much simpler at
the symmetry axes because all first derivatives vanish:
$\partial_\theta \mu(r,\theta)= \partial_\theta \phi(r,\theta) =
\partial_\theta J=0$; the result on the minor axis turns out to be not 
very useful in providing constraints to models. We remark that, e.g.,
circular orbits in an axisymmetric system trivially satisfy eq.\
\myref{Heq} because $q_\mu=q_\phi=\lambda=1$.

\begin{figure} 
\epsfxsize=8.4truecm 
\centerline{\epsfbox{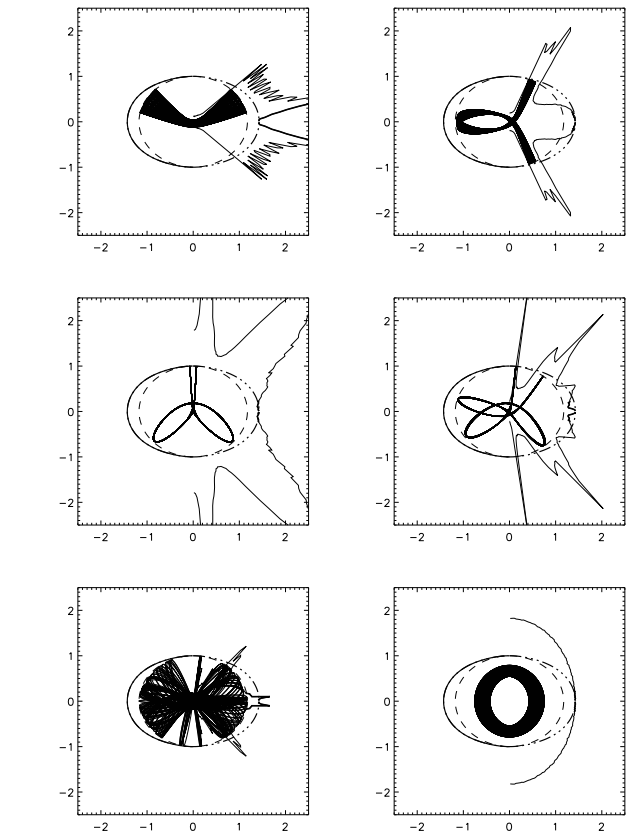}}
\caption{As Figure 1, but for an elliptic disc with a surface density 
power-law slope $\gamma =1-\alpha=1.2$ and axis ratio $q=0.7$. Only
the fish orbit supports the potential near the major
axis.}\label{a-2q7.ps}
\end{figure}

\section{Orbits in elongated discs}

Consider scale-free discs with surface density of the form: 
\beq\label{ellip}
\mu(r,\theta) \propto r^{-\gamma}  
\left(\cos^n \theta + q^n\sin^n\theta\right)^{-{\gamma \over n}}. 
\eeq
The gravitational potential of these discs is not exactly elliptical,
and is formally given by eq.\ \myref{pot}. It can be computed with
harmonic expansions as in SZ98. When $n=2$, the surface density is
stratified on ellipses of axis ratio $q$.

Figures~\ref{a2q7.ps} and~\ref{a-2q7.ps} show the density distribution
of the dominant orbits in two such discs. The heavy solid line in each
panel indicates the shape of the orbit after all the scaled copy
orbits and their reflection images are summed up in a scale-free
fashion; for clarity only the right hand part is drawn.  We find that
near the major axis the density contributions of all tube orbits, and
nearly all boxlet orbits, curve in the opposite sense as the model
density; the exceptions are the fish orbits in some potentials
(Miralda--Escud\'e \& Schwarzschild 1989; K93), which sometimes
support the curvature of the model density.

\begin{figure} 
\epsfxsize=10.0truecm 
\centerline{\null\hskip3.0truecm \epsfbox{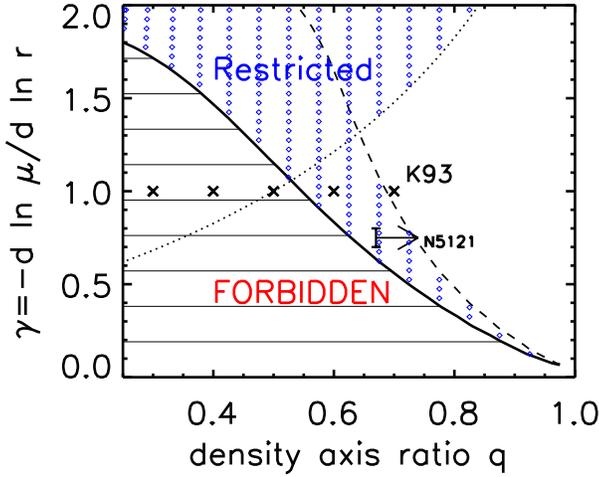}}
\caption{The parameter space for scale-free discs: cusp-slope $\gamma$ 
versus axis ratio $q$. No self-consistent elliptic discs can occupy
the shaded region to the left of the solid line (where $1-q>
\gamma/2$), due to the wrong curvature of orbits at the major axis.
Regions to the left of the dashed curve ($h<0.5$, see \S 5.2) and/or
the dotted line ($C<0.707$, see \S 5.3) are also restricted by the
shapes of the available orbits. The crosses indicate logarithmic
elliptic discs which are found to be likely non-self-consistent in
previous numerical studies by Kuijken (1993).  The error bar indicates
the cusp strength for the nuclear bar in NGC5121 observed by
HST.\looseness-2 }
\label{nuker.ps}
\end{figure}
 
The above result suggests that a large fraction of scale-free elliptic
disc potentials cannot be made self-consistent near the major axis.
While the tube orbits are always anti-aligned with the potential
(e.g., de Zeeuw, Hunter \& Schwarzschild 1987), once the `backbone' of
a cored potential, namely the set of box orbits, is replaced by the
boxlets and stochastic orbits of a cusped model, there may be no
orbits left to support the potential near the major axis.  Since a
general property of all orbits is that the amplitude of the orbital
angular momentum always reaches a local maximum on the major axis, an
orbit tends to spend more time away from the major axis than on the
axis.  As a result, the curvature of orbits is often opposite to the
real density even after all `cousin' orbits are added, which limits
the range of non-axisymmetric models with cusps that can be
self-consistent.

\section{Fitting the curvature at the major axis} 

We now present a more quantitative analysis of the curvature of the
regular orbits.  A simple and necessary condition for self-consistency
near the major axis of the model follows from eq.\ \myref{Heq}: at least
some orbits should have a negative $\left[q^{-2}_\phi-(1+\gamma -\gamma
\lambda)\right]$ and others a positive value in order to place the average 
at zero.  This places a range on the possible curvature (or axis
ratio) of the potential, as follows:
\beq\label{kp}
[0, 1+\gamma -\gamma \lambda_{\rm max}]_{\rm max} \le q^{-2}_\phi \le 
[1, 1+\gamma -\gamma \lambda_{\rm min}]_{\rm min}.
\eeq
Since $\infty > \lambda_{\rm max} \ge \lambda\ge \lambda_{\rm min} >0$
by definition (cf eq.\ \myref{lambda}), a clean non-trivial limit is
that
\beq\label{gamma}
q_\phi > {1 \over \sqrt{1+\gamma}}.
\eeq
This condition turns out to be quite powerful. It implies that discs
with shallow density cusps (small $\gamma$) cannot have a highly
elliptical potential because they would violate the requirement of the
curvature. The curvature condition is trivially satisfied in the
Keplerian regime, since the potential becomes spherical ($q^{-2}_\phi
\rightarrow 1$).  This would be the case near a central black hole, or
at very large distance of a model with a finite mass.

\begin{figure} 
\epsfxsize=10truecm 
\centerline{\null\hskip-0.5truecm\epsfbox{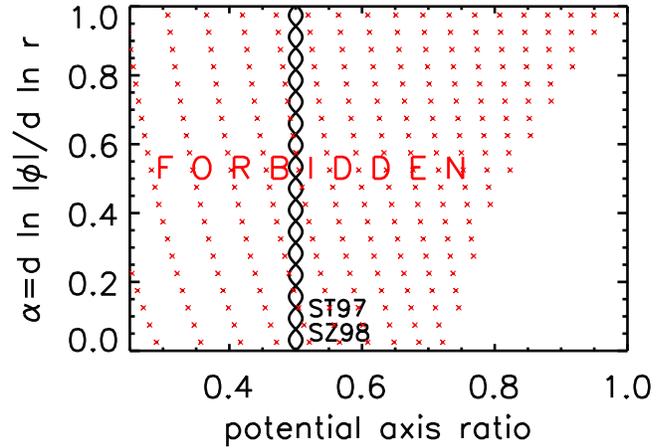}}
\caption{The parameter space of non-axisymmetric disc potentials.
Self-consistency is ruled out for a large fraction of discs with
elliptical density contours (indicated by small crosses) and for the
whole sequence of cuspy, separable potentials of axis ratio $0.5$
studied by Sridhar \& Touma (1997) and Syer \& Zhao (1998)
(figure-eight symbols).  The axis ratio here is for the equal
potential contours, although they are generally not exactly
elliptical. }
\label{sz.ps}
\end{figure}

Our analysis nicely confirms the result of the semi-analytical study
of SZ98 that all ST97 discs are non-self-consistent. Eq.\
\myref{gamma} is always violated for these discs, because
\beq
q_\phi = {1 \over \sqrt{3 - \gamma}}< {1 \over \sqrt{1+\gamma}} \quad 
       \mbox{\rm for ST97 discs with $\gamma<1$;}
\eeq
ST97 models with $1<\gamma<2$ have unphysical negative density regions.

More interesting is the application of the curvature criterion
\myref{gamma} to the entire class of elongated discs. Self-consistent
elliptic discs ($n=2$ in eq.\ \myref{ellip}) must have strong enough
cusps such that
\beq
\gamma > q^{-2}_\phi-1 \approx 2(1-q_\mu),
\eeq
where $q_\mu=q$ is the axis ratio of the surface density.

Figure~\ref{nuker.ps} shows that about 50\% of the parameter space in
the cusp strength vs axis-ratio plane of elliptic discs can be simply
ruled out on the basis of examining the curvature at the major axis.
Figure~\ref{sz.ps} shows a similar diagram, but for the potential of
elliptic discs and the ST97 models. When applied to the specific case
of the $\gamma=1$ elliptic disk, the lower limit on the allowed axis
ratio provided by eq.\ \myref{gamma} is in harmony with the numerical
estimates of K93.

Perhaps unexpectedly, our criterion shows that non-axisymmetric discs
with a shallow cusp are easier to rule out than those with strong
cusps, while the opposite has been suggested for three-dimensional
models, based on numerical experiments with two classes of potentials,
one with a finite force at the center ($\gamma=0$ in projection), the
other with the force diverge as $r^{-1}$ ($\gamma=1$) (Merritt \&
Fridman 1996).  These two opposing suggestions here may well reflect
the still incomplete coverage of parameter spaces of both approaches:
we are restricted to 2D scale-free discs, but we gain better coverage
of the range of cusp slope $\gamma$ and ellipticity $1-q$ owing to our
analytical approach.

\section{Discussion and conclusions}

We have shown that many scale-free elliptic discs cannot be
self-consistent, and we have given specific criteria for picking out
non-self-consistent models.  The major axis is a local maximum in
terms of the angular momentum of an orbit, which often translates to a
local minimum in terms of the fraction of time spent by regular
boxlets and tube orbits. This is opposite to that required by the
surface density distribution. A clean result from this analysis is
shown by the forbidden zone in Fig.~\ref{nuker.ps}, which implies that
galactic nuclear discs with a shallow density profile ($0 \le \gamma
\le 0.3$) are necessarily nearly axisymmetric.

\subsection{Hidden assumptions}

Our results apply only to orbits which cross the disc major axis with
a finite angular momentum $\|J\| \equiv r^2 \|\dot{\theta}\|$, so that
$\|\dot{\theta}\|^{-1}$ and its derivatives are well-defined at
$\theta =\pi/ 2$.  This condition clearly breaks down for the axial
orbits and/or the stochastic orbits.  The two are perhaps the same
since the axial orbits are destabilized by the divergent force at the
center, and become stochastic.

Stochastic orbits make our method problematic because they could reach
$J=0$ virtually anywhere. And it is quite inevitable that they will be
populated during galaxy formation.  However, the way they are
populated must be restricted in an equilibrium model because
stochastic orbits are slowly-evolving orbits. The only sure recipe to
mix stochastic orbits of the same energy ($E_0$) together into a
time-independent building block (a super-orbit) is with a distribution
function $\delta(E-E_0)$ (Zhao 1996); this distribution mixes in the
regular orbits of the same energy $E_0$ as a side effect. In our
scale-free models these super-orbits form one family, with the
relative weights of all members related by a simple scaling (\S 2).
Fortunately for the present problem, the spatial distribution of this
super-orbit family is always as round as the equal-potential contours.
Since the equal-density contours are generally more elongated,
including the super-orbits does not relax our constraints on
curvature.\footnote {It is possible to modify the shape of a
super-orbit by subtracting the densities of regular orbits from
it---always keeping the density everywhere non-negative. It is not
clear whether the resulting component (a nearly spherical distribution
with many worm-holes) will be more elongated than the density
model. Even if it were to do so, we suspect that it would not relax
the constraint on the intrinsic shape of the model unless this single
component is also very heavily populated during galaxy formation.}  So
in summary, our result remains valid if stochastic and/or axial orbits
are populated with a time-independent distribution.

Can a small operation near the major axis of the density model bring
the model to self-consistency?  One possibility is to modify the shape
of the equal-density contour near the major axis by a significant
amount (without violating the Poisson equation and the positivity of
the density) such that it is rounder than that of the equal potential
contour ($q_\mu > q_\phi$) {\it locally}.  It is conceivable that by
heavily populating the super-orbit component, which is stratified on a
set of mildly elongated equal-potential contours, one might counteract
the wrong curvature from regular orbits at the major axis.  Whether
self-consistency can be restored this way also depends on the maximum
fraction of mass which can be allocated to the super-orbit component.

\subsection{Expanding the forbidden territory} 

Can self-consistent scale-free discs be ruled out on the right-hand
side of the `forbidden zone' in Figure~\ref{nuker.ps}? Perhaps this is
likely: if regular orbits always have a finite (rescaled) angular
momentum $h$ when crossing the major axis, and orbits with less
angular momentum are stochastic, then the minimum angular momentum
sets up a barrier to prevent any regular orbit from falling to the
center or touching the zero-velocity curve on the major axis.  Here
$h$ is defined by $0<h\equiv J^2_{\rm min} / J^2_{E} \le 1$, where
$J_{E}$ is the maximum angular momentum allowed for an orbit of energy
$E$.  There should be a gap in phase space filled by stochastic orbits
which separates the unstable periodic axial orbit from the regular
boxlets.  Such a stochastic gap is typically seen in the start space
on the zero-velocity surface of three-dimensional models, and in
surfaces of section along the major axis of two-dimensional models
(e.g., fig.\ 5 of Schwarzschild 1993).  The minimum angular momentum
sets up a dynamical boundary to $r$, $r\partial_r\phi$, $\dot{r}$ and
$\lambda$ (cf eq.\ \myref{lambda}) at the upper and lower ends.  The
upper limit on $\lambda$ typically helps little to tighten the
constraints, but the finite lower limit of $\lambda$, namely
$\lambda_{\rm min}(h)$ as a function of $h$, propagates to a more
stringent upper limit on the flattening (cf eq.\ \myref{kp}) and
pushes the limit of the forbidden region further to the right in
Figure~\ref{nuker.ps}.  The parameter space to the left of the dashed
line can be ruled out as long as the stochastic gap is wide enough
($h>0.5$). Typically only fish orbits or higher-order resonant orbits
can have a small $h$, and we suspect that it is difficult to reach
self-consistency for much of the region to the right of the forbidden
zone in Fig.~\ref{nuker.ps} without heavily populating the
higher-order resonances.  This hypothesis is also supported by the
range of non-self-consistent models of K93 (indicated by the crosses),
which extends to our $h=0.5$ line.

\subsection{Constraint from the density contrast}

The density ratio of the minor vs.\ major axis should also set limits
on the ellipticity of self-consistent discs.  As shown by SZ98, if
\beq
C \equiv \left[{\mu(r,0) \over \mu(r,{\pi \over 2})}\right] 
  \left[{\phi(r,0) \over \phi(r,{\pi \over 2})}\right]^{\gamma \over \alpha} 
\equiv \left({1 -\epsilon_\mu \over 1-\epsilon_\phi}\right)^{\gamma},
\eeq
is defined as a measure for the model density contrast between the
minor and major axes, where $1-\epsilon_\mu$ and $1-\epsilon_\phi$ are
the minor-to-major axis ratio of the equal density and equal potential
contours respectively, then $C$ must have a non-trivial lower and
upper bound,
\beq
C_{min} \le C \le C_{max},
\eeq
set by the shape of orbits in the potential.  A model constructed by
populating only the thinnest banana orbit in a potential cannot be
flatter than a certain value because of the fact that even the
thinnest banana orbits spend a fair amount of time near the minor
axis.  Populating thick boxlet orbits, higher-order resonant orbits
and tube orbits tend to make the shape rounder as these orbits come
more frequently to the minor axis than the thin banana orbits.
Figure~\ref{nuker.ps} also shows the curve with $C=0.707 \approx 1 /
\sqrt{2}$ in the $q$ vs.\ $\gamma$ plane.  This curve is interesting
for reference as it corresponds to a configuration where only the
thinnest banana orbits in a ST97 potential are populated.  In contrast
a $C=1$ curve (which means $\epsilon_\mu=\epsilon_\phi$) would
correspond to a configuration where only the $f(E)$ super-orbits are
populated.  The realistic configurations are likely in between these
extremes.  Combined with the criteria from the curvature,
Figure~\ref{nuker.ps} suggests that about $2/3$ of the parameter space
of cusped elliptic discs are ruled out.

\subsection{Implications for nuclear discs and bars}

Our curvature constraint on the existence of self-consistent
scale-free elliptic disks can be applied to the properties of observed
stellar bars and nuclear discs.  The dynamics of these highly
flattened systems are essentially two-dimensional with the small
vertical oscillation completely decoupled from the motion in the
plane.  In the absence of a curvature criterion for three-dimensional
models, it is premature to discuss the parameter space for elliptical
galaxies.

Tumbling stellar bars have scales set by the corotation radius.
Numerical experiments show that regular boxlet or loop orbits have a
central `hole' of finite size compared to their apocenter.  So one
expects a region infinitely close to the center where no `large'
orbits will ever visit. The dynamics of the very nucleus will be
dominated by `small' orbits {\it in situ}.  Such a situation would not
be possible for models with a finite core.  We know of at least one
set of strongly elongated bar models --- the rotating Freeman (1966)
elliptic discs, which can be built self-consistently for any axis
ratio.  Our curvature condition clearly does not apply to these models
with an analytical core ($\gamma=0$) as the finite force at the center
stabilises axial orbits and box orbits. But it can be applied with
confidence to the central regions of cusped elliptical nuclear
bars. And our results show that these cannot be strongly elongated;
their axis ratio in projected light should satisfy the major axis
curvature condition eq. \myref{gamma}.

Our curvature constraint on elongation comes from the general property
that the orbital angular momentum peaks near the major axis, so that
the orbital density generally has a local minimum on this axis,
whereas the model density is largest there.  We therefore expect it to
hold to some extent for general discs, as long as a genuine cusp
($\gamma \ne 0$) in the central surface density creates a divergent
force at the center and destabilises any box orbit.  An outer
truncation of the mass distribution and the tumbling motion of the bar
can greatly change the potential at large radius, but at very small
radii the model reduces to the static self-gravitating scale-free case.
This happens at radii well-inside the ILR of any rotating potential
(but well beyond the sphere of influence of any small central black
hole if it exists in bars) such that the centrifugal and Coriolis
forces are negligible compared to the self-gravity of the cusp.

Measurements of nuclear cusp slopes for galaxies with nuclear bars are
as yet scarce. The one object for which the light distribution has
been measured with HST resolution, NGC 5121, shows a steep light
profile with $\gamma=(0.75\pm 0.05)$ and a projected axis ratio
$(0.67\pm 0.05)$ (Carollo \& Stiavelli 1998).  This data point, as
shown in Fig.~\ref{nuker.ps}, is safely outside the forbidden zone
with or without correcting for the inclination. Imaging of other
nuclear bars are needed to establish whether there is a zone of
avoidance for the shapes of cusped galaxies.

\medskip
\noindent
It is a pleasure to acknowledge helpful discussions with Vincent Icke,
Konrad Kuijken, Doug Richstone, Massimo Stiavelli, Scott Tremaine, and
Dave Syer, and thoughtful suggestions by an anonymous referee which
helped to improve the presentation of our results.

\appendix
\section{Derivation for double derivative of $\Gamma$}

\noindent
The lengthy calculation of the double derivative of $\Gamma$ with
respect to $\theta$ is greatly simplified near the major axis
($\theta={\pi \over 2}$), where we can safely ignore any first or
third derivatives of an even function with respect to time $t$ or
angle $\theta$, since these tend to zero.

First we decompose the double derivative of $\Gamma$
(cf. eq.~\ref{basic}) at $\theta={\pi \over 2}$ to that with respect
to radius $r$, to angular momentum $J$, and to the angular
part $s(\theta)$ of the density,
\beq
{\d^2 \Gamma \over \Gamma \d\theta^2}\vert_{\theta={\pi \over 2}} = 
-{\d^2 J \over J \d\theta^2} 
+{\d^2 r^\gamma \over r^\gamma \d\theta^2}
-{\d^2 s(\theta) \over s \d\theta^2}.
\eeq
With the help of the equations of motion (eq.\ \ref{eom}) we find
\bey
{\d^2 J \over J \d\theta^2} 
&=&{\d \dot J \over J \dot \theta \d\theta} \\
&=&-{ \partial^2_\theta \phi \over  r^2 \dot \theta^2}\\
{\d^2 r^\gamma \over r^\gamma \d\theta^2}
&=&\gamma {\ddot r \over r \dot \theta^2} + 
   \gamma (1+\gamma)\left({\d r \over r\d\theta}\right)^2\\
&=& \gamma { {\dot \theta}^2 r - \partial_r \phi 
 \over r \dot \theta^2} 
+ \gamma (1+\gamma)\left({\dot r \over r\dot \theta}\right)^2.
\eey
Now rewrite the second derivatives of the density $\mu$ and the
potential $\phi$ in terms of the curvatures of the density and
potential, which by definition (cf. eq.\ \ref{qdef}) satisfy
\bey\label{aqmu}
q_\mu^{-2} - 1 &=& {\partial^2_\theta \mu \over r \partial_r \mu}
= {\partial^2_\theta s(\theta) \over s} {1 \over -\gamma},
\\\label{aqphi}
q_\phi^{-2}- 1 &=& {\partial^2_\theta \phi \over r \partial_r \phi}
= {\partial^2_\theta p(\theta) \over p} {1 \over 1-\gamma}.
\eey
Rewriting ${\dot r}^2$ and ${\dot \theta}^2$ in terms of $K_r$ and
$K_\theta$, we obtain
\beq
{\d^2 \Gamma \over \Gamma \d\theta^2} \vert_{\theta={\pi \over 2}}
= K_\theta^{-1} \left[
q_\phi^{-2}-(1+\gamma)(1-\gamma K_r) + \gamma q_\mu^{-2}K_\theta 
\right],
\eeq
which immediately reduces to eq.\ \myref{Heq}.

{}

\label{lastpage}
\end{document}